\documentclass{aipproc}

\usepackage{epsfig}
\usepackage{natbib}

\layoutstyle{8x11double}

\begin{document}

\title{NUMERICAL METHOD FOR SHOCK FRONT HUGONIOT STATES}

\author{J. M. D. Lane}{
  address={Center for Nonlinear Dynamics, University of Texas at Austin},
  email={mlane@chaos.ph.utexas.edu}
}

\author{M.  Marder}{
  address={Center for Nonlinear Dynamics, University of Texas at Austin},
  email={marder@chaos.ph.utexas.edu}
}


\copyrightyear  {2005}

\begin{abstract}
We describe a Continuous Hugoniot Method for the efficient simulation
of shock wave fronts.  This approach achieves significantly
improved efficiency when the generation of a tightly spaced
collection of individual 
steady-state shock front states is desired, and allows for the study of 
shocks as a function of a continuous shock strength parameter, $v_p$.  This 
is, to our knowledge, the first attempt to map the Hugoniot continuously.  
We apply the method to shock waves in Lennard-Jonesium along the
$<$100$>$ direction.  We obtain very good agreement with prior
simulations, as well as our own benchmark comparison runs.
\end{abstract}

\date{\today}

\maketitle

\section{Introduction}
On-going experiments the University of Texas at Austin are investigating
the shock-strength dependence of melt time scales in tin using terawatt 
laser systems on sub-picosecond time scales. Large numbers of trials exploring 
many different shock velocities will be possible. We present here a method 
intended to provide direct comparisons with these experiments.

There are two major difficulties with applying traditional simulation
methods \cite{germann.00} \cite{kadau.04} \cite{holian.98} to the study of the dynamics near the shock front:  (1) To
produce the environment at the front, one must simulate a large and
ever-growing system, of which the front constitutes only a very small
fraction; and (2) The conditions within a steady-state shock take long
times to arise, and each computationally-expensive shock run results
in only a single data point. 

The constrained dynamics methods of the Hugoniostat and others
\cite{maillet.00} \cite{reed.03} offer a solution to
the first issue, but provide no information about non-equilibrium
dynamics at the shock front that would be needed to compare with
experiments.  The approach of Zhakhovskii et
al. \cite{zhakhovskii.99} succeeds in addressing the first point at
the shock front, but does not address the second.  We generalize and
expand on their methods.  First, we concentrate our efforts on
the neighborhood of the shock interface, thereby increasing
computational efficiency, and second, we map system response to a
continuum of shock strength final states in a single run.  These, combined, constitute the Continuous
Hugoniot Method.

\section{Continuous Hugoniot Method}
\begin{figure}[htb]
\epsfig{file=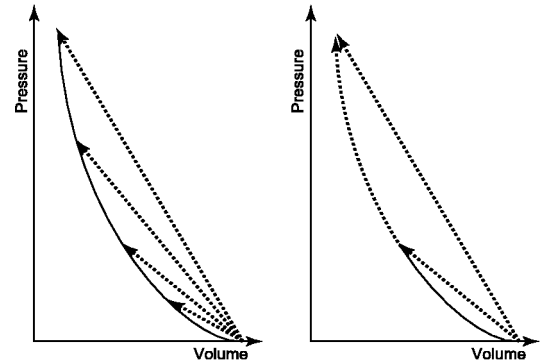,width=2.75in}
\caption{(left) The Hugoniot, as a collection of final shock states; and (right) as the state-to-state path of the Continuous Hugoniot Method.}
\label{f:ourpath_hugoniot}
\end{figure}

Our simulation method makes the Hugoniot the thermodynamic path of our
simulation system.  This has not been possible experimentally.  Figure
\ref{f:ourpath_hugoniot} contrasts the experimental loading paths
(Rayleigh lines) to each state of the Hugoniot and the loading
path which we will use to reach the same state points. 

We outline our simulation method in four stages:

\noindent \underline{Benchmark} -- We begin each simulation with a
traditional shock wave computation.  This initial long duration,
full-system run is allowed to continue to its final shock steady
state.  This steady state is our first point on the shock Hugoniot and
serves to seed our subsequent computations.  In Figure
\ref{f:ourpath_hugoniot} this is represented by the lower Rayleigh
line from the initial state to the beginning of the Hugoniot ramp. 

\noindent \underline{Reduce} -- We truncate the system
to a fixed width neighborhood around the front by removing particles 
which are beyond a set distance behind the shock front.  We impose a 
warm impactor (piston) boundary condition and match the thermodynamic 
statistics at the rear purge point using a strong Langevin thermostat.  
A buffer of pristine material is preserved ahead. The length of the 
reduced system is determined by the system's thermalization length.  
The piston driving velocity and mean velocity of the thermostat are 
equal, and this value, $v_p$, serves as the external control parameter 
for shock strength. 

\noindent \underline{Shock strength ramp} -- To this reduced system,
we introduce a quasistatic increase of the shock forcing parameter.
The goal is to increase the shock strength, while always maintaining a
direct mechanical coupling between the forcing at the piston and the
response at the shock interface.  The piston velocity increases by a
set amount per timestep.  The temperature of the stochastic forcing is
updated every purge to match the distribution at the purge point.  The
point of forcing always remains a set distance behind the shock front.
This process continues until the shock strength parameter has reached
its terminal value.

\noindent \underline{Terminal benchmark comparison} -- Finally, a second
benchmark run is made with a shock strength matched to the final value of
the shock strength ramp.  This allows the final state of the Hugoniot
ramp to be compared directly to a traditional shock run to identify
any problems and to serve as an error check. 

\vspace{0.1in}
\noindent {\bf Validity of Shock States} \\
The validity of our purge technique assumes that the back of our system 
is in equilibrium.  We verify this by velocity distribution analysis 
before we reduce the system.  We assume our shocks are always in
steady state.  This is guaranteed, so long as our 
ramp loading is quasistatic.  If so, the small changes have time to 
equilibrate across the entire system and the driving and response are 
mechanically coupled.  If not, then the foundation of the Hugoniot-Rankine 
equations is eroded, and off-Hugoniot states are produced.

The essential property of this method is our ability to very
slowly ramp the shock strength parameter while maintaining correct
thermodynamic conditions at a roughly constant distance behind the
front. This is not currently achievable in experiment. Instead, in
experiments driving forces are imposed at ever-increasing distances.
If one ramps the driving velocity in an experimental situation,
the result is isentropic compression rather than a shock
response.

\vspace{0.1in}
\noindent {\bf Velocity Ramp Rate} \\
We can estimate an upper bound for the quasistatic ramp rate of the shock 
strength control parameter.  The velocity is scaled by units of the wave 
speed of the compressed material, $C_S$, and the time is scaled by the return time of the acoustic waves in the system, $2L/C_S$.  We, thus, get the nondimensional 
condition for a quasistatic ramp.
\begin{equation}
\dot{\tilde{v}}_p = \frac{d \tilde{v}_p}{d \tilde{t}} =  \frac{d (v_p/C_S)}{d (t/\frac{2L}{C_S})} \ll 1 \quad \Rightarrow \quad \frac{dv_p}{dt} \ll \frac{C_S^2}{2L}
\label{e:ramp_rate}
\end{equation}
Note that the upper bound for quasistatic
ramp rate goes to zero for large systems.  The advantage of the method is lost 
when systems are too large.

\section{Application to Lennard-Jones}

As a prelude to more realistic but computationally intensive studies, we test these ideas with the Lennard-Jones
potential, which has a well-documented solid shock response \cite{holian.95} \cite{holian.98}.

\vspace{0.1in}
\noindent{\bf Simulation Details} \\
We use the cubic-spline Lennard-Jones 6\,--\,12 potential \cite{holian.91} in order to allow easy comparison with published Hugoniot results of Germann et al. \cite{germann.00}.  The shock was oriented along the $<$100$>$ direction of the fcc crystal with unit cell dimension $5.314$~\AA~$= 1.561 \sigma$.  Initial temperature was varied with a weak Langevin thermostat from zero to $10 {\mathrm K} = 0.083\,\,k_B{\mathrm T}/\epsilon$.  Results are found not to depend on the initial temperature for shock driving velocities, $v_p$ above $0.75\,C_o$.  Systems were 20 $\times$ 20 lattice planes in cross-section with transverse periodic boundary conditions.  The timestep was 0.3 femtoseconds.

The traditional shock simulation runs (benchmark runs) were driven by a
warm impactor and reached $600$~\AA~in length ($\sim$\,100,000 particles).
Continuous Hugoniot Method runs were held to $200$~\AA~in length
($\sim$\,20,000 particles), at any one time.  The cumulative distance 
covered by these treadmilling runs was almost 1.3
$\mu$m in length, and would have required over 1,200,000 particles in
a conventional simulation.  Every shock was given time to
establish a steady state (usually 30 to 60 ps).  The ramp rate of $0.001\,\mbox{m/s \,\, per \,\, step} = 3.3 \times 10^{12}\,\mbox{m/s}^2$ is used for the remainder of this article.  We report on our efforts in the strong shock regime, for driving velocities ranging from $v_p = 0.75\,C_o$ to $1.5\,C_o$.  

\begin{figure}[htb]
\epsfig{file=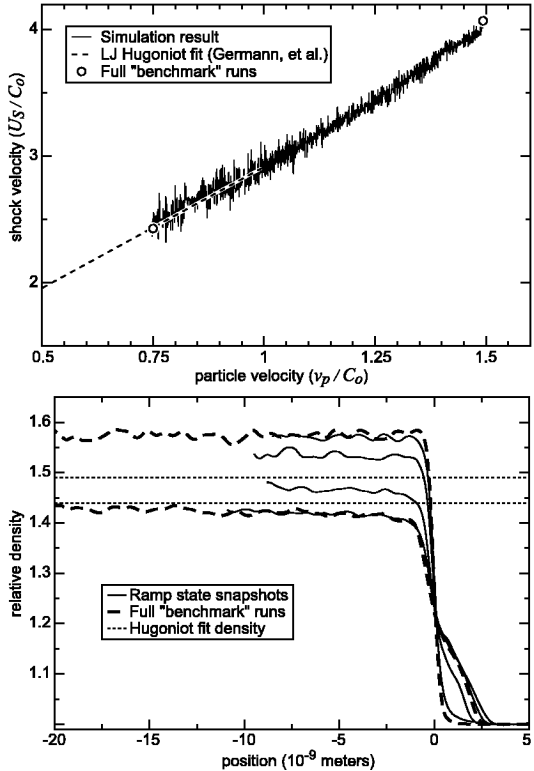,width=2.75in}
\caption{(top) Results of the Continuous Hugoniot Method, the fit of Germann et al. \cite{germann.00}, and bookend benchmark runs.  (bottom) Density profiles snapshots from a shock strength ramp (solid), from the initial and terminal benchmarks (dashed), and from the Hugoniot fit (dotted).}
\label{f:LJ_profile_comparison}
\end{figure}

\vspace{0.1in}
\noindent {\bf Principal Hugoniot Results} \\
The Continuous Hugoniot Method allows a
system to move directly from one shock state to another.  Therefore,
the path of our Continuous Hugoniot Method through $U_s$--$v_p$ space
during a single run is the principal Hugoniot of final
shock states in the material.  This is true to the extent that there is 
quick convergence within the reduced system to the values
far behind the shock.  Figure \ref{f:LJ_profile_comparison} shows such a path
for a simulation which runs from $v_p=0.75\,C_o$ continuously through $v_p=1.5\,C_o$.
Initial and terminal benchmark runs, bookend the ramp.  The Hugoniot fit proposed by
Germann et al. is plotted in its applicable range.  

We see excellent agreement of our method's results with both
comparisons.  In the lower range of $v_p$, where we can compare to the
published fit, our data overlays the fit very nicely and 
continue it smoothly beyond the range for which it
was originally published.  At higher shock strengths our 
data stiffens (as it should), showing a
super-linear increase in $U_S$ vs $v_p$.  In the upper range of $v_p$,
we compare to our terminal benchmark results, which also agree well. 

\vspace{0.1in}
\noindent {\bf Comparison of Density Profiles} \\
Figure \ref{f:LJ_profile_comparison} shows a series of four density
profile snapshots taken from a continuous Hugoniot ramp (shown as solid lines).
They are at $v_p=0.75\,\,C_o$, $0.9\,\,C_o$,  $1.2\,\,C_o$ and $1.5\,\,C_o$.
The density profiles of the two benchmark runs and the final densities predicted
by the Hugoniot fit for $0.75\,\,C_o$ and $0.9\,\,C_o$  are plotted.

The profiles produced by the Continuous 
Hugoniot Method agree well with the results of the benchmark runs up to 
the point where they are purged.  The benchmark densities, however, 
continue to grow beyond this point.  In both cases the density predicted by 
the published fit is approximately 2\% larger than the average final density 
produced by  the Continuous Hugoniot Method.  It appears that the 
density requires a larger spatial region than we
have provided in order to converge completely to its asymptotic value.

\begin{figure}[htb]
\epsfig{file=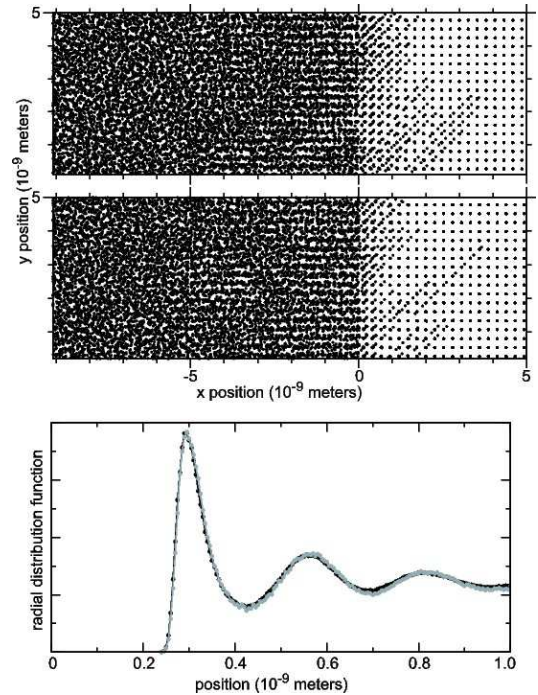,width=2.75in}
\caption{{\bf Structural dynamics and rdf comparison} -- Particle slices in
  the region of the shock front for states from (middle) the
  Continuous Hugoniot Method and (top) the terminal benchmark.  Also shown (bottom)
  is a comparison of radial distribution functions for the compressed material 
  extending 10 nm behind each front.  $v_p = 1.5\,\,C_o$.}
\label{f:LJ_particle_comp_2000}
\end{figure}

\vspace{0.1in}
\noindent {\bf Comparison of Final States} \\
The final state of the Hugoniot ramp is a particularly important point of 
comparison because it is the state of the maximum integrated error.
Figure \ref{f:LJ_particle_comp_2000} provides snapshots from the
terminal benchmark (top) and the Continuous Hugoniot Method (bottom).
The slices are along the $x$--$z$ plane at the final piston velocity
$v_p=1.5\,C_o$.  Both systems exhibit a strong disordering
transition on similar length scales, islands of incomplete
disordering, and a common sharp density rise.  Both have also
developed forward-reaching features ahead of the front.  Figure
\ref{f:LJ_particle_comp_2000}, shows very good agreement between the 
radial distribution functions for the material behind each front.

\vspace{0.1in}
\noindent {\bf Efficiency and speed up} \\
The computational speedup provided by the Continuous Hugoniot Method
depends upon the density of points with which one wants to locate the
Hugoniot.  We find that the computation time to compute  two 
benchmark runs by traditional methods is approximately equal to the 
computation time we employed to compute the entire intermediate Hugoniot, via our
method. Thus if we had chosen to trace out the Hugoniot by
interpolation with 10 conventional shock computations, we would
have required 5 times more computation. 
More generally, let  $N_h$ be the number of traditional runs needed
to map the Hugoniot between two states as a function of shock strength.  Then
the speed up is linear in $N_h$, given roughly by $N_h/2$.
Alternatively, we point out we were able 
to simulate the cumulative effect of 1,280,000 particles with the resources
required to hold only 20,000 at any one time.

\section{Conclusions}
We have presented the Continuous Hugoniot Method for efficient
simulation of the dynamics in steady-state shock fronts.  We confirmed
in a Lennard-Jones system
that the loading path of the Continuous Hugoniot Method followed the published
Hugoniot fit.  Comparison of particle snapshots and radial
distribution functions at the final state of the Continuous Hugoniot
Method ramp also showed good agreement with traditional shock
methods. 

All these measurements were made with greatly reduced
computational expenditure over traditional methods.  These savings
are proving critical as the method is applied to more realistic and
computationally costly potentials such as tin. 

\begin{theacknowledgments}
We thank Todd Ditmire, Stephan Bless and Will Grigsby
for useful conversations. This work was supported by the National
Science Foundation under DMR-0401766, and DMR-0101030 and by the U.S. Dept. of Energy, National Nuclear Security Administration under Contract DE-FC52-03NA00156.
\end{theacknowledgments}

\bibliographystyle{aipproc}
\bibliography{../bibtex/JMDL.bib}

\begin{thebibliography}{7}
\expandafter\ifx\csname natexlab\endcsname\relax\def\natexlab#1{#1}\fi
\providecommand{\enquote}[1]{``#1''}
\expandafter\ifx\csname url\endcsname\relax
  \def\url#1{\texttt{#1}}\fi
\expandafter\ifx\csname urlprefix\endcsname\relax\def\urlprefix{URL }\fi

\bibitem[Germann et~al.(2000)]{germann.00}
Germann, T.~C., et al., \emph{Phys. Rev.
  Lett.}, \textbf{84}, 5351--5354 (2000).

\bibitem[Kadau et~al. (2004)]{kadau.04}
Kadau, K., et al., \emph{Int. J. of Modern Phys. C},
  \textbf{15}, 193-201 (2004).

\bibitem[Holian and Lomdahl(1998)]{holian.98}
Holian, B.~L., and Lomdahl, P.~S., \emph{Science}, \textbf{280}, 2085--2088
  (1998).

\bibitem[Maillet et~al.(2000)]{maillet.00}
Maillet, J.~B., et al.,
  \emph{Phys. Rev. E}, \textbf{63},
  16121 (2000).

\bibitem[Reed et~al.(2003)]{reed.03}
Reed, E.~J., et al., \emph{Phys. Rev.
  Lett.}, \textbf{90}, 235503 (2003).

\bibitem[Zhakhovskii et~al.(1999)]{zhakhovskii.99}
Zhakhovskii, V.~V., et al.,
  \emph{Phys. Rev. Lett.}, \textbf{83}, 1175--1178 (1999).

\bibitem[Holian(1995)]{holian.95}
Holian, B.~L., \emph{Shock Waves}, \textbf{5}, 149--157 (1995).

\bibitem[Holian et~al.(1991)]{holian.91}
Holian, B.~L., et al.,
  \emph{Phys. Rev. A}, \textbf{43}, 2655--2661 (1991).

\end{thebibliography}

\end{document}